# Plasma-assisted Discharges and Charging in EUV-induced Plasma

Mark van de Kerkhof[a,b*], Andrei M. Yakunin[a], Vladimir Kvon[a], Selwyn Cats[a], Luuk Heijmans[a], Manis Chaudhuri[a], Dmitry Asthakov[c]

a) ASML Netherlands B.V., De Run 6501, 5504 DR Veldhoven, The Netherlands
b) Department of Applied Physics, Eindhoven University of Technology, Eindhoven, The Netherlands
c) ISTEQ B.V., Eindhoven, The Netherlands; and Institute for Spectroscopy of the Russian Academy of Sciences, Troitsk, Moscow, Russia

**Abstract.** In the past years, EUV lithography scanner systems have entered High-Volume Manufacturing for state-of-the-art Integrated Circuits (IC), with critical dimensions down to 10 nm. This technology uses 13.5 nm EUV radiation, which is transmitted through a near-vacuum $H_2$ background gas, imaging the pattern of a reticle onto a wafer. The energetic EUV photons excite the background gas into a low-density $H_2$ plasma. The resulting plasma will locally change the near-vacuum into a conducting medium, and can charge floating surfaces and particles, also away from the direct EUV beam. This paper will discuss the interaction between EUV-induced plasma and electrostatics, by modeling and experiments. We show that the EUV-induced plasma can trigger discharges well below the classical Paschen limit. Furthermore, we demonstrate the charging effect of the EUV plasma on both particles and surfaces. Uncontrolled, this can lead to unacceptably high voltages on the reticle backside and the generation and transport of particles. We demonstrate a special unloading sequence to use the EUV-induced plasma to actively solve the charging and defectivity challenges.

*****Mark van de Kerkhof,** E-mail: mark.van.de.kerkhof-msd@asml.com

1. ## Introduction

The ongoing technological evolution in Integrated Circuits (IC's) is driven by an exponential growth in demand for computing power and data transport, and is expected to accelerate further in coming years with the advent of Artificial Intelligence running partly on high-performance centralized servers but also on local and mobile edge-computing devices. Power consumption and computing performance will be key drivers for improving architectures as well as further increases in pattern density. Driven by Moore's law[1], named after Intel co-founder Gordon Moore, the critical dimensions of IC's have shrunk by a factor of 2 every 1.5-2 years; nowadays, the critical dimensions of the most advanced devices are in order of 10 nm. This has been enabled by advances in all processing steps, but mainly by continuous advances in photolithography, by decreasing the (UV) wavelength and increasing the numerical aperture of the photolithographic tools (also known as scanners – see fig. 1), and introducing resolution enhancements such as polarization[2,3] and immersion[4]. Recently, the introduction of Extreme Ultra-Violet (EUV) scanners into high-volume manufacturing[5] has ensured that Moore's law can continue for the coming years[6].

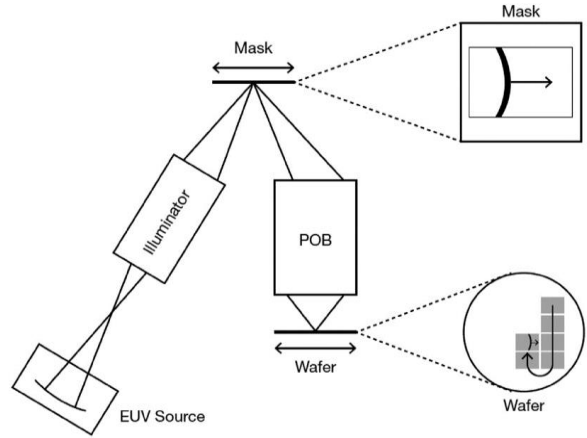

Figure 1: Basic principle of an EUV scanner; the object on a reticle (or mask) is illuminated and imaged onto a portion of a wafer by the Projection Optics Box (POB) while scanning; after which the wafer is moved to a new position and the process is repeated (source: ASML).

Even with the outstanding imaging and overlay capability of the EUV scanners, device yield can still be affected adversely by other factors, such as particles ending up on critical surfaces. The current state-of-the-art lithography node has critical device structures of below 20 nm at wafer level. This translates to 80 nm at reticle level (multiplied by 4x, because of the scanner de-magnification M=1/4), and a printing particle size limit of $d_c$=~40 nm at reticle level. For future nodes, this printing defect size will decrease further. Therefore particle contamination control, or defectivity control, is a key aspect of scanner system design. The industry is pursuing a dual-path approach to defectivity. One path is advanced particle contamination control to ensure zero particles reaching the sensitive reticle and wafer surfaces. In parallel, EUV-compatible pellicles are now available (transparent films shielding the reticle from particles), which are fully supported by the scanner[7]. Pellicles have intrinsic benefits in terms of particles, but come at the cost of reduced system transmission and productivity; the cost/risk-driven trade-off between the two options depends on application details and can vary for different chip manufacturers and even for different exposure layers[8]. This paper will focus on the case of advanced particle contamination control, without pellicle.

Particles can come from parts and scanner integration, or can be generated by the moving parts in the scanner or can be carried in with the reticles and wafers. It was found that the EUV-induced plasma in the scanner can be a major factor in releasing and transporting particles via electrostatic release and under-etching[9]. This plasma is the result of ionization of the protective hydrogen background gas in the scanner[10]. Besides direct impact on particles, the EUV-induced plasma can also interact with electrostatics in several ways: it can e.g. reduce the safe voltage in terms of gas breakdown. In the subsequent sections these aspects will be described in more detail and design consideration will be discussed.

2. **EUV-induced plasma**

Current EUV Sources for lithography are based on EUV emission by a hot Sn (tin) plasma: this Sn plasma is formed by irradiating a stream of Sn droplets by a high-power pulsed IR laser (LPP – laser-produced plasma)[11]. The Sn plasma has a peak temperature of several 10's of eV, which efficiently emits EUV radiation around 13.5 nm. High conversion efficiency is achieved by an IR pre-pulse to enlarge the target and reduce the Sn density[12]. The raw emission spectrum is broadband, and includes longer-wavelength Vacuum UV (VUV) components, but in the scanner this spectrum is filtered by the narrow-band Bragg-reflection mirrors to 13.5 nm (± 0.2 nm)[13]. Current LPP EUV Sources are highly transient, firing short <100 ns pulses with energy of ~5 mJ at 50 kHz (250W output), with the peak of EUV in the first 50 ns and a tail of broadband radiation.

EUV lithography employs a low-pressure background gas of 1-10 Pa hydrogen ($H_2$), to maintain self-cleaning conditions for the sensitive EUV mirrors in the optical system of the scanner. Hydrogen was chosen as background gas, because of high chemical activity of H-ions[14], high transmission of $H_2$ for EUV, and low/negligible sputtering by the light H-ions. The 92 eV EUV photons will lead to some absorption and photo-ionization of the hydrogen background gas, creating a plasma, as outlined in fig. 2. Because of the low absorption of $H_2$ (attenuation coefficient α = 0.0078 m$^{-1}$ at 5 Pa), the ionization degree will be low at ~10$^{-4}$ % (for a 250 W Source), and secondary interactions between hot electrons and ions will have low probability. Momentum conservation dictates that the large excess energy of 76 eV is carried by the photoelectron while the ions remain at room temperature[15]. These hot photoelectrons will lose energy by secondary ionizations and dissociations of the neutral hydrogen molecules within first 2 μs; this results in an increase of plasma density even after the EUV pulse has passed, and to formation of up to 3 pairs of ions and electrons per absorbed photon[16].

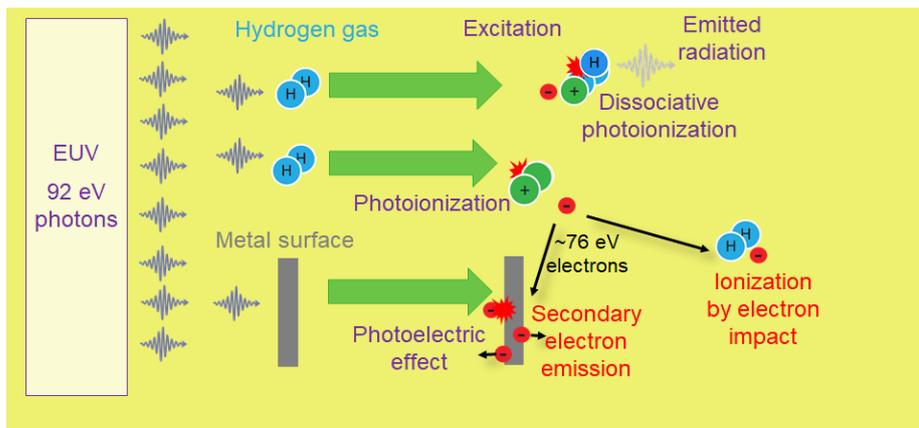

Figure 2: Basic processes of EUV-induced hydrogen plasma.

When the electron energy is reduced to beneath 20 eV by these secondary events, subsequent cooling will slow down and the plasma will decay by diffusion and wall recombination. In view of the low ionization degree and the proximity of wall surfaces, volume recombination will be a minor effect, and can be effectively ignored[17]. During the EUV pulse of <100 ns, the plasma will instantaneously expand several cm's beyond the EUV beam by the fast photo-electrons, and subsequently expand and decay by diffusion. Given the LPP Source frequency of 50 kHz, this cycle repeats every 20 μs; as the decay time of the plasma will typically exceed 20 μs at pressures of 1 - 10 Pa[18], there will be build-up of a steady-state plasma, with repeating highly transient peaks every 20 μs, as illustrated in fig. 3. The resulting quasi-steady-state background plasma can be described as weakly coupled, diffuse and cold; but the transient plasma will typically not be in local thermal equilibrium (LTE), and the electron energy distribution will not be Maxwellian[19]. This in turn means that many classical plasma assumptions will not or not always apply and care must be taken with classical equations for plasma temperature, Debye shielding length and plasma sheath.

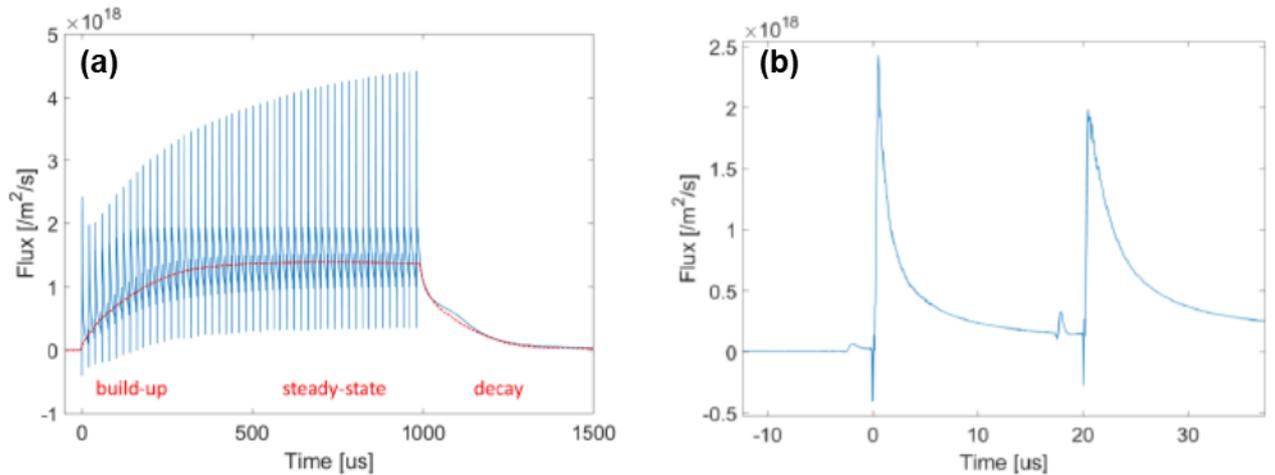

Figure 3: Left: Build-up, steady-state and decay of pulsed EUV-induced plasma, as measured by Retarding Field Energy Analyzer (RFEA) in scanner-like test-stand directly after LPP Source exit[20]. Right: zoom-in on first two pulses; also visible is the minor plasma formation due to the pre-pulse.

The energetic photoelectrons will set up a significant plasma-to-wall potential difference in order of 76 V, peaking during the EUV pulse and dropping fast in the afterglow[21]. In practice, this plasma-wall potential will be reduced by photo-electric effect which results in low-energy electrons being released from any wall (or mirror) irradiated by the EUV beam or by EUV flare[22], by secondary electron emission, and by charge compensation by ions; the resulting steady state potential of floating surfaces and dielectrics exposed to the EUV-induced plasma is in order of a few volts (typically ~2 V). As the mean free path of the energetic photoelectrons can be several cm's (given electron-neutral collision cross-section $\sigma$ of ~$10^{-20}$ m$^2$ at ~76 eV; from Tawara[23]), the plasma can be significantly larger in dimensions than the EUV beam itself, and also charging of floating surfaces to negative potential can occur up to significant distances within the vessel.

The plasma sheath is the potential drop region near the wall surface, and will depend on the local plasma density and electron temperature[24]. Even though care should be taken with using standard equations for sheath thickness and potential drop over the sheath, since the underlying thermal equilibrium assumptions are not always satisfied, they give a good approximation for the steady-state background plasma, which is likely to dominate the average behavior over time. For a 250 W Source, the sheath thickness can be estimated to be in order of ~0.1-1 mm, increasing to several mm's away from the EUV beam; the electric field at the surface can be estimated to be up to ~10 kV/m close to the beam, and decreasing sharply away from the beam.

For particle contamination control, the zone around the reticle is of specific interest, as particles on the reticle have the most severe impact[25]. The reticle is clamped to a scanning stage and faces downwards, with metal reticle masking blades and other conductive surfaces in close proximity. The resulting slits suppress plasma diffusion of plasma, and transport of plasma through these slits is largely driven by the fast photo-electrons, as shown in fig. 4. It should be noted that the EUV-induced plasma will be different for different locations in the scanner, since every successive mirror in the optical system will absorb ~30% of light[26]; still, as the reticle is in the center of the scanner system, it can be taken as a reasonable first approximation for the entire scanner.

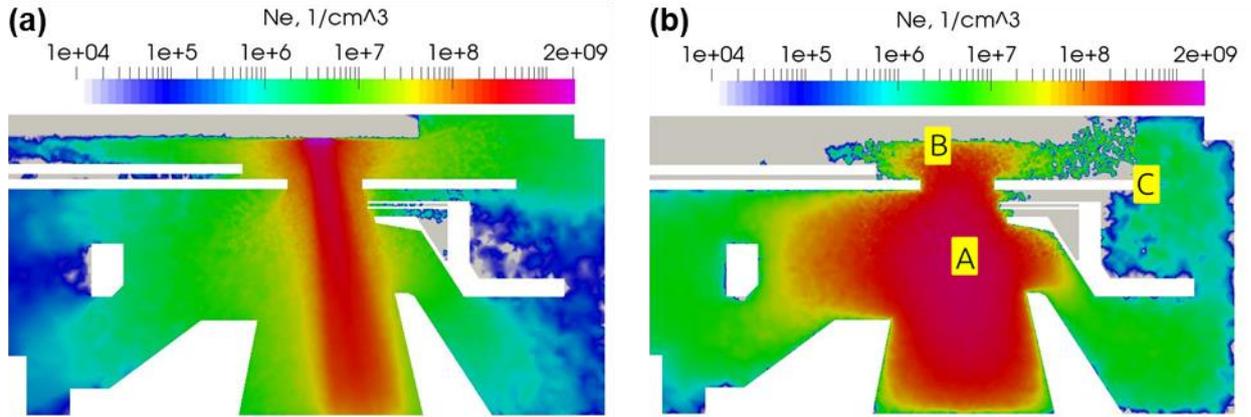

Figure 4: Schematic of reticle zone, showing EUV beam region (A), floating reticle surface (B), grounded reticle masking blades (C). Left (a): during isolated EUV pulse the electrons from the EUV beam penetrate through the slits. Right (b): plasma accumulation beyond the beam confines over multiple pulses.

At first glance these plasma parameters seem indicative of a glow discharge, but the pulsed photo-ionization origin of the plasma leads to important differences, such as strong transients and non-Maxwellian energy distribution function during and after the EUV pulse (of <100 ns). This precludes the use of fluid-like models, which rely on continuity equations for moments of the distribution functions for electron density, velocities and energies. Instead, a kinetic model must be used that can solve the full equations for the electron distribution functions without any a priori assumptions about their shapes, such as (Monte-Carlo) Particle-in-Cell (PIC)[27]. The essence of this model consists of a Poisson equation solver, followed by updating the charged particles positions and velocities based on the obtained electric field distribution and individual particle velocities. This model has been tailored for simulation of EUV-induced plasma[28], and validated at a relevant pressure of 5 Pa in an off-line test setup using an Electrostatic Quadrupole Plasma (EQP) Analyzer[29]. Recently, we have further extended this model to a full 3D PIC model, with options to speed up calculations by hybridization of the model with fluid-like model for the cooled electrons.

3. **Plasma and electronics: plasma-assisted discharges**

Classically, the risk of discharges is described by the Paschen criterion. This describes the condition where the gain factor by cathode electron generation plus gas ionizations by accelerating electrons exceeds the loss factor of electrons to the anode surface, to trigger a self-amplifying discharge[30]. This requires that electrons can gain enough energy between collisions, requiring a sufficiently long mean free path and high enough electric field, but also have sufficient collisions. These considerations yield a voltage threshold as function of gas type, pressure and distance, above which a self-amplifying avalanche effect will occur, driving the current through the gas sharply up. This can be plotted as breakdown voltage versus pressure times distance (p.d), which is called the Paschen curve (see fig. 5). Compared to air, hydrogen has a relatively low minimum breakdown voltage of 273 V, at p.d = 1.5 Pa.m[31]; for pressures of ~5 Pa this translates to critical distances of ~30 cm.

For a near-vacuum system, left of the minimum, the Paschen criterion in principle allows very high voltages but care should be taken for long discharge path lines (e.g. to vessel walls), and for points of field amplification, such as a sharp edge or protrusion, or a particle, especially at the anode[32]. A notable concern is coating edges, such as on both backside and frontside of the reticle coating edge, which will have an effective submicron edge radius, resulting in significant field amplification (which can acerbated by the triple point junction of dielectric glass substrate, conductive coating and vacuum). Also, care should be taken that AC or switching voltages can reduce the Paschen threshold[33]. Given the high energy densities, a Paschen-discharge may easily both generate and release particles, mainly by local overheating at the point of contact of the electrons.

While the Paschen criterion has proven to work well in ambient conditions (to the right of the minimum), care should be taken in (near-)vacuum, for several reasons: surface properties and feedback mechanisms become more important w.r.t. gas properties, adsorbed gases can become dominant over background gas (especially $H_2O$, but also e.g. $O_2$ and $N_2$), and curved electrical field lines at electrode edges can lead to longer discharge paths[34]. At the same time, surfaces can act as electron sinks; it has been shown that the minimum breakdown voltage is somewhat increased in presence of surfaces, as shown in fig. 5, but the steep slope for the low-p.d regime is significantly reduced suppressed: even at low p.d values, critical discharge voltages remain well limited to well below 1 kV, in clear contradiction of Paschen prediction[35].

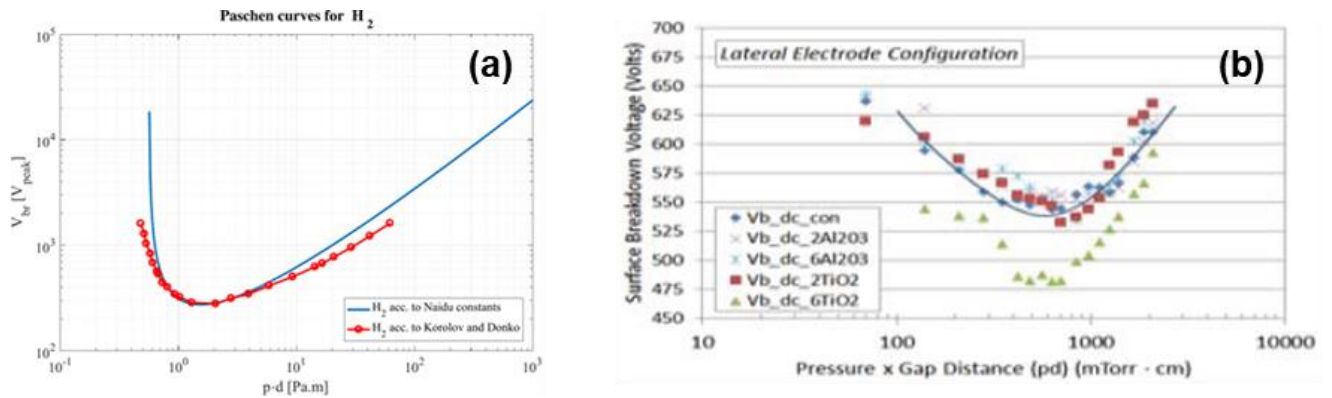

Figure 5: Left (a): Paschen curve for $H_2$, showing good match between analytical expression and measurements[36]. Right (b): Modified breakdown curves in presence of a surface (insulating epoxy, with different nanoparticle additions), showing a more gentle slope at low p.d-values (reproduced from Li[37]).

In-house experiments confirmed these trends. A slowly increasing voltage (2 V/s) was applied to a reticle placed on top of a standard baseplate in a low-pressure $N_2$ environment, with insulating polyimide spacers of ~100 μm. This confirmed discharges at voltages well below the Paschen prediction and with a weaker pressure dependence than predicted by classical Paschen theory (fig. 6), consistent with the findings in fig. 5. Figure 6 also shows that while lower pressure allows for somewhat higher voltage, the amplitude of the discharge is larger. Tests at 5 Pa did not show discharges up to 800 V.

The discharges caused particle removal from the reticle. In the experiment, the baseplate was seeded with 5 μm $SiO_2$ particles. These were observed to be removed from the baseplate and transferred to the reticle. This is not simple electrostatic release, as one might expected for field strengths in order of ~5 MV/m[38], because it is not observed for 5 Pa and lower pressures, which were exposed to the highest fields: up to 8 MV/m. It is correlated to the local discharges that happen at pressures of 10 and 40 Pa at lower field strengths (respectively 7 MV/m and 4.5 MV/m). Our proposed explanation is that the ~5 μm particles act as field amplification points to trigger discharges which also release the particle, but this hypothesis could not be positively confirmed due to insufficient accuracy in measuring pre and post conditions.

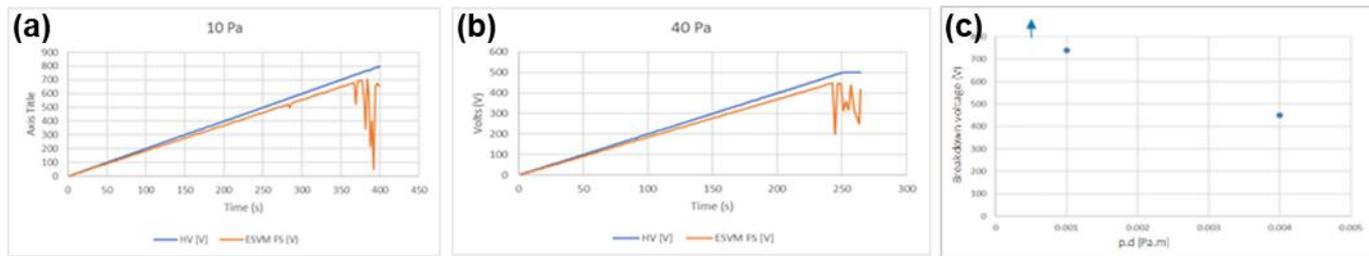

Figure 6: Left (a) and middle (b): gradual voltage ramp at 2 V/s until breakdown for 10 and 40 Pa; blue line is voltage to frontside, red line is frontside voltage as measured by ESVM. Right (c): summary of the breakdown voltages for different p.d values (no discharge was observed at 5 Pa for up to 800 V maximum)[39].

When surface aspects dominate, this also implies a higher likelihood of surface flashovers versus through-gas discharges to opposite surfaces[40]. For reticle discharges as outlined above, this is a concern for frontside defectivity, since a flashover from the charged backside via the floating frontside to the (grounded) baseplate would increase the risk of particles being generated that can reach the frontside. In view of this, it is advised to have dissipative reticle support studs to allow the reticle to gradually de-charge to the baseplate.

In presence of plasma, the free electrons and ions invalidate the basic Paschen assumptions, and result in a significant shift of the avalanche threshold to lower voltages. In itself, the plasma will not focus either electrons or ions, so local discharge-like damage such as overheating is not to be expected from plasma. However, in combination with an external voltage, current focusing can indeed occur, and such a plasma-assisted discharge can induce surface damage and create particles. This is obviously a concern for the high-voltage electrostatic clamps used in the EUV scanner[41], so these must be perfectly shielded from EUV, including the volume extending several cm's around the EUV beam. Less straightforward is that this is also a concern for switching power supplies and circuit boards for fast sensors, which in practice can have voltages above 100 V[42], so these also must be properly shielded from the EUV-induced plasma.

Interaction of a plasma with biased electrodes can lead to formation of different structures[43]. When the bias voltage of positively biased electrode becomes too high, formation of the so-called "fireball" structure may occur. Formation of a "fireball" in the scanner vessel is not intended, as it is very similar to a discharge, and a large current can be focused into a small area. In the presence of slits and complicated geometries the conditions for formation of the "fireball" and similar discharges will be different as compared to bulk plasma.
We studied this with our PIC model, and validated experimentally in a set-up with a simplified geometry, as shown in fig. 7. This geometry was modeled using a fast 2D PIC model, for computational efficiency, with the same underlying physics and cross sections as our 3D-PIC model used for more realistic geometries.

The model shows a breakdown or discharge towards the positive anode when the plasma is switched on. This can be explained by electrons being accelerated towards the anode and achieving sufficient energy for further ionizations of hydrogen molecules. As the electrons are accelerated further to the positive anode, a positively charged plasma cloud is formed which screens the electrode potential and moves the zone of electron acceleration away from the electrode, thus effectively forming a channel of current, as illustrated in fig. 8. This is a similar mechanism to streamer formation in tip-shaped anodes[44]. Once this channel is fully formed, after ~2-3 µs, breakdown is complete as shown by the sharp increase in current.

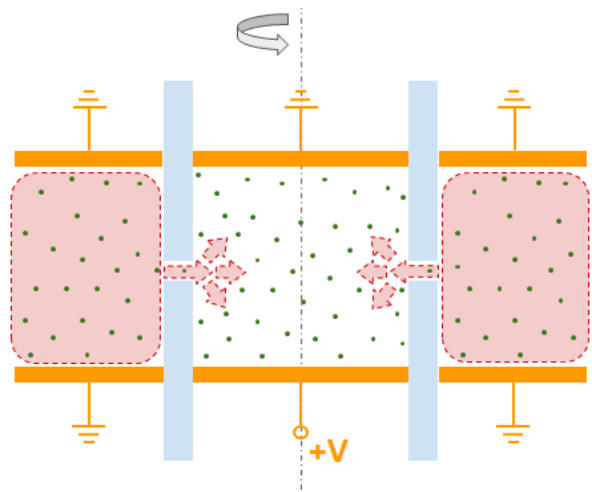

Figure 7: Simplified 2D model geometry. Depicted are the cylindrical tube with a dielectric wall (gray), electrodes (orange) and plasma-filled region (pink). Neutral H2 gas is represented with green dots.

In contrast to Paschen theory, the plasma-assisted breakdown is not determined by the product of distance and pressure: for a given distance, higher pressure and/or higher plasma power result in higher plasma electron density and lower breakdown voltage.

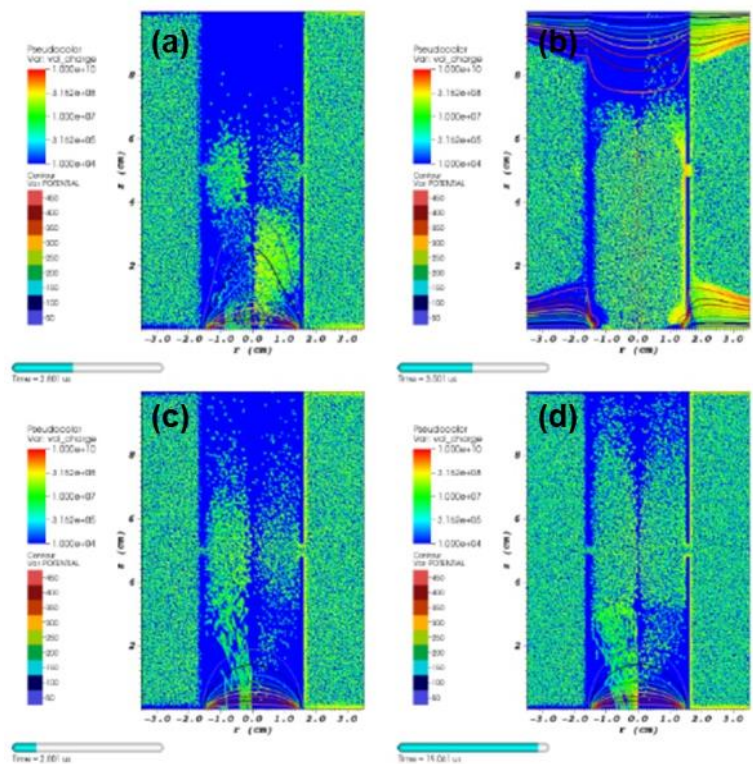

Figure 8: Simulated plasma dynamics for 10 Pa (a,b) and 1 Pa (c,d), showing electron and ion densities (left and right side of each image) and voltage contour lines, for specific time stamps. Anode potential was set to 500 V. The snapshots for 10 Pa show the transition to breakdown at ~3 µs, while the snapshots for 1 Pa show the more or less stable anode glow.

As fig. 9 shows, the model shows a peaked threshold behavior in the current at the moment of breakdown for higher pressure (10 Pa); this is the "Fireball" mode. For lower pressure (1 Pa), the model shows an oscillating current, but no breakdown, since the number of ions formed in this case is too low to achieve sufficient screening of the positive electrode to move the acceleration zone away from the anode and form the conductive channel as above. In that case, the plasma formed will remain in an oscillatory anode glow mode.

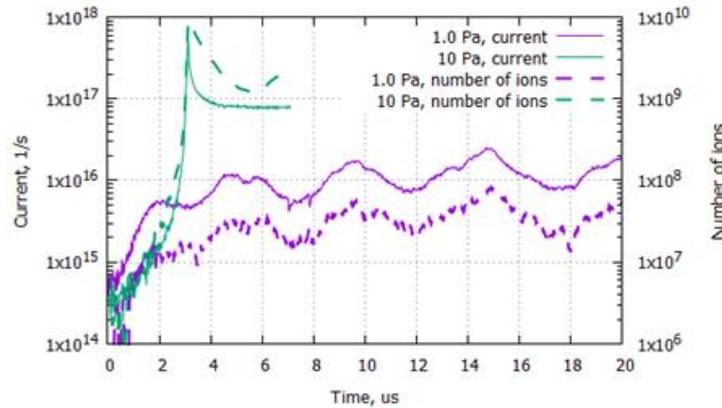

Figure 9: Model showing avalanche and breakdown at 10 Pa, with sharply peaked increase in current (green line); and oscillating glow discharge at 1 Pa (purple line).

The experimental set-up is essentially a cylindrical tube with two electrodes and a possibility to add free charge carriers from an RF plasma, as shown schematically in fig. 10. Hydrogen pressure was varied in the range of 1 – 10 Pa, and RF power varied between 10 to 60 W and the distance between electrodes was varied in range of 1 – 10 cm. After applying a given combination of these parameters, the bias voltage on the positive electrode is scanned from 0 – 250 V, remaining always below the Paschen minimum of hydrogen, while the electrode currents was measured continuously by a Keithley 2010 multimeter.

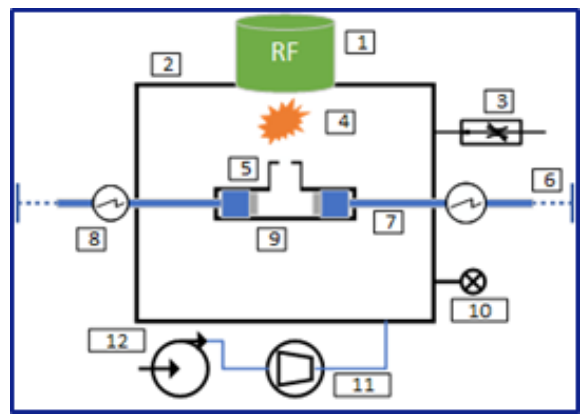

Figure 10: Experimental setup for plasma-assisted discharge: 1 – Cesar 1312 RF generator; 2 – stainless steel chamber; 3 – hydrogen valve; 4 - hydrogen plasma; 5 – glass tube; 6 - micrometer translation stage; 7 – insulated PTFE rod with a connector; 8 – power supply/multimeter; 9 – stainless steel electrode; 10 – pressure gauge; 11 – turbo-pump; 12 – rotary pump.

The resulting breakdown threshold was observed to be in order of ~100 V, significantly below the predicted Paschen threshold for this configuration and even well below the theoretical Paschen minimum of 273 V. It was also confirmed that the avalanche is directed towards the anode, while the ion current to the cathode is ~10x lower. As predicted by the model, the moment of breakdown shows a sharp peak in current, after which a steady high current flows and voltage drops somewhat, as shown in fig. 11. Varying conditions of electrode distance and pressure resulted in observations of breakdown below and around the Paschen minimum voltage, even for low pressure and short distances, in what classically should be a 'safe' zone of p.d-V combinations.

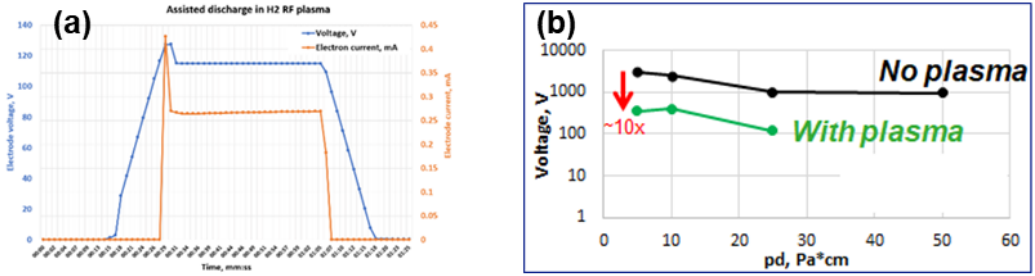

Figure 11: Left (a): example of plasma-assisted discharge and current when ramping up voltage between electrodes while plasma switched on in neighboring chamber (using 60W plasma, 5 Pa, 10 cm distance). Right (b): examples of significant reduction of breakdown voltage when plasma is switched on (10W plasma; 5 cm distance; 1, 2 and 5 Pa); both with and without plasma, the observed breakdown voltages at low pressures are all significantly reduced with respect to classical Paschen prediction of >$10^5$ V (see fig. 6), which is attributed to the glass tube wall surfaces.

In general, the critical ion density $n_i^{crit}$ to trigger a plasma-assisted discharge can be estimated from the condition of positive electrode screening: the potential drop due to volume charge should be comparable with the electrode potential drop:

$$n_i(t) > n_i^{crit} \sim \frac{2\epsilon_0}{e} \cdot \frac{\varphi}{h^2} \sim 10^8 \; cm^{-3} \qquad (1)$$

With the electrode potential drop $\varphi \approx$ 30-70 eV (to accelerate electrons to energies at which ionization is most efficient[45]), the region of ion accumulation $h \approx$ 1 cm, and $e$ the elementary charge. This estimate of $n_i^{crit}$ is in line with the 2D PIC simulations above. In more complicated geometries, the exact value of the critical ion density will depend on the (in)homogeneity of the electric field and the distance of the plasma source to the anode. Still, rather than a discharge threshold being determined by voltage and the product of pressure and distance, as in the case of Paschen, the discharge threshold is now driven by applied voltage $V_{ext}$, pressure $p_{H2}$ and local plasma density, which in turn scales with pressure and EUV power $I_{EUV}$ and with a suppression factor $\gamma_{supp}$ that describes the fall-off of plasma away from the EUV-beam.

$$n_i^{crit} \sim V_{ext} \cdot p_{H2}^2 \cdot I_{EUV} \cdot \gamma_{supp} \qquad (2)$$

Equation 2 illustrates that the risk of plasma-assisted discharge needs to be re-evaluated for any increase in either EUV power, local pressures or external voltages. For the complicated internal geometries of an EUV-scanner, no analytical expression can be used. However, our 3D PIC model, with the same underlying physics and cross sections as the 2D PIC model as used and validated above, can now be used to check any design proposal for safe limits on local voltages and pressures plasma-assisted discharges.
Also, as general guidelines, floating or insulating surfaces should be avoided as much as possible to minimize the risk of surface flashovers, and edges and protrusions should be sufficiently rounded to avoid dangerous field amplification points.

## 4. reticle charging and discharges

As outlined in section II, floating surfaces and dielectrics close to the EUV-induced plasma may become charged. In particular the reticle needs to be considered in this respect, since it consists of an insulating glass substrate with a conductive coating on the backside (for electrostatic clamping purposes), and a conductive reflective multi-layer coating on the frontside; both conductive layers are floating independently; this is outlined in fig. 12.

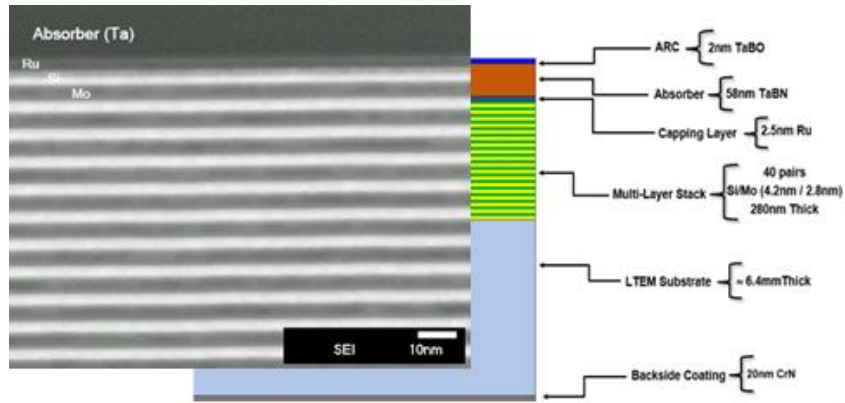

Figure 12: Sketch of reticle cross-section, showing conductive coating stacks on frontside and backside; adapted from McLellan[46]

The reticle backside is clamped electrostatically to a movable positioning module, while the frontside is directly exposed to the EUV beam and EUV-induced plasma. Grounding of the reticle is impractical in view of the risk of particles generated when making electrical connection through the oxide top layers on the moving/scanning reticle[47]. During exposures the reticle frontside will acquire a transient potential due to competing direct photoelectric effect from EUV irradiation (driving to positive) and subsequently de-charging from plasma, and will return to ~0 V after every pulse, as shown in fig. 13.

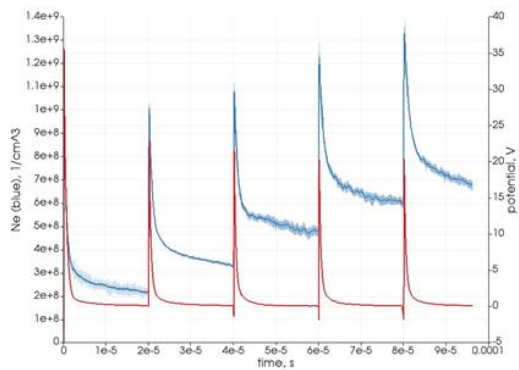

Figure 13: PIC simulation of electron density (blue line) and reticle frontside surface potential (red line), showing accumulation of plasma over pulses but no build-up of surface potential.

The reticle backside is quite different, since this is not exposed directly to EUV irradiation and shielded by the clamp. The backside coating plane is connected to the plasma volume through only a small gap, which acts as a spatial filter to suppress diffusion for both positive and negative charges. Still, two effects can result in charging of the reticle backside: and secondary gas ionizations and secondary electron emission by the surrounding clamp, of which secondary gas ionizations are expected to be dominant.

Secondary ionizations in the gas surrounding the reticle by the energetic photoelectrons result in electrons propagating more or less isotropically around the actual EUV beam, which allows electrons to reach the conductive backside coating even if the coating is recessed from the edge. The ions have a lower likelihood to reach the backside as these are accelerated more along the electrical field lines, scatter less and have higher inertia; so ions will likely hit surrounding surfaces and stick there. This results in a net negative charging of the reticle backside, as is shown schematically in figure 14.

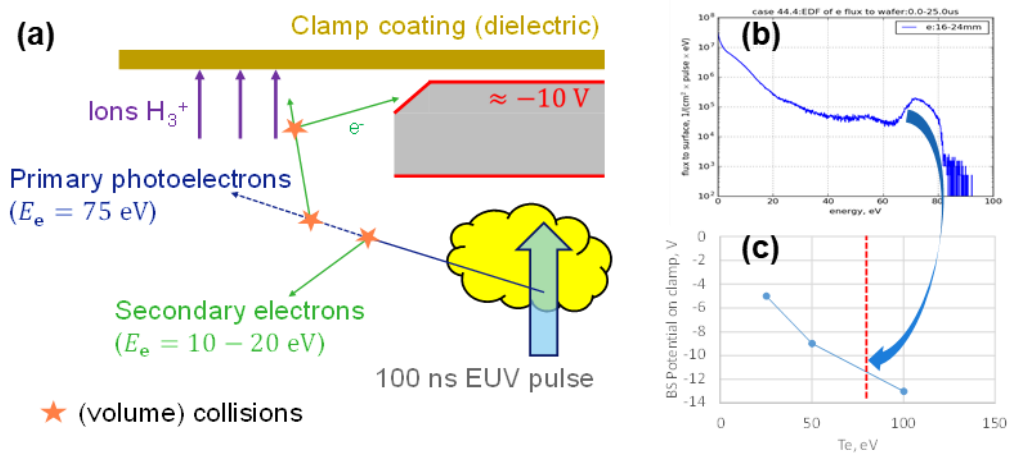

**Figure 14: Left (a): basics of charging mechanism of reticle backside; Top right (b): energy spectrum of electrons reaching reticle edge modelled by 3D PIC code. Bottom right (c): energetic secondary electrons charge reticle backside, while ions are carried by field and momentum towards nearest wall resulting in a net negative charge on reticle.**

As the reticle backside charge and voltage build up, electrons will be repulsed and ions attracted, which will result in an equilibrium charge and voltage, which will depend on details of plasma (e.g. EUV power, pressure and beam position with respect to reticle edge). PIC modelling for the reticle geometry in NXE:3400 of electron spectrum reaching the backside reticle edge shows that the cumulative process of charging by fast electrons and partial neutralization by ions result in an equilibrium negative voltage in order of -10 V. This voltage, and the associated excess electrons, will remain on the reticle backside after the plasma fully decays at the end of exposures. The backside voltage has been simulated to scale inversely quadratically with increasing pressure, as shown in fig. 15; this is due to the combined effect of reduced average frontside potential and increased collisions at higher pressure which act to reduce the high-energy tail of the EEDF.

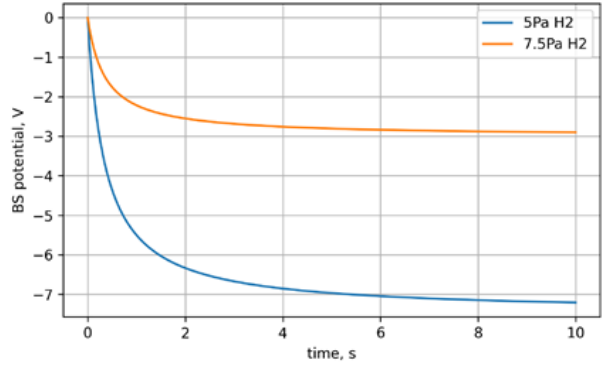

Figure 15: PIC simulations of backside voltage dependence on time and pressure, showing >2x reduction in voltage for 1.5x higher pressure.

Also reticle geometry and coating details are relevant: a recessed coating will show a higher equilibrium charge since this will increase the spatial filtering of ions, and more so than for electrons (due to secondary electron emission or bouncing of electrons from surfaces). This implies an additional consideration for the reticle backside coating beyond the existing specifications for clamping, with the coating preferably extended as close as possible to the edge. The acquired backside voltage may seem negligible, but during reticle unloading the backside voltage is amplified by the changing capacitance between reticle and clamp, as outlined in fig. 17, while the charge locked onto the floating surface remains constant[48]. This is reflected in the basic equations:

$$U = \frac{Q}{C} = \frac{Q \cdot d}{\epsilon_0 \cdot A_{cl}} \qquad (3)$$

$$U_{unl} = U_{cl} \cdot \frac{d_{unl}}{d_{cl}} \qquad (4)$$

With $A_{cl}$ the (constant) area of the clamping electrode, $U_{unl}$ the backside potential during unload, $U_{cl}$ the backside potential as clamped, $d_{cl}$ the distance as clamped, which is in order of microns, and $d_{unl}$ the distance during unloading, which is in order of cm's (as sketched in fig. 16), giving a potentially ~1000x increase.

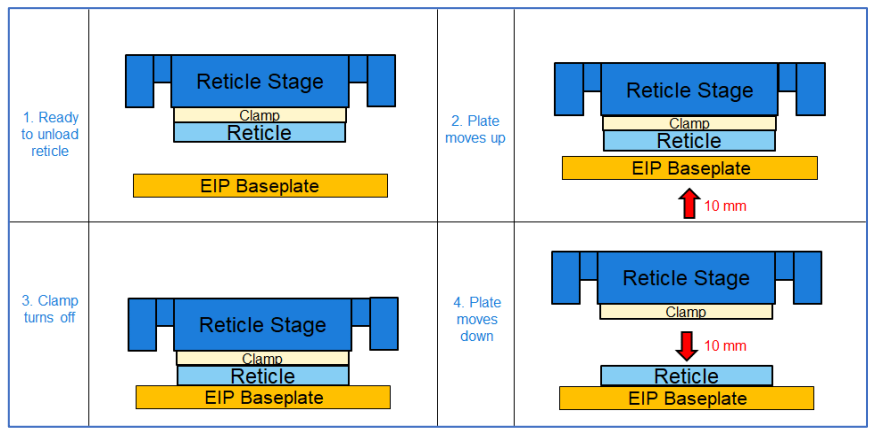

Figure 16: Illustration of reticle unloading sequence. While clamped the separation between reticle backside and clamp electrode is in order of a few micrometers, resulting in strong capacitive coupling; during unload the separation is increased to ~1 cm, reducing the capacitance by several orders of magnitude.

In reality, capacitive coupling between backside and frontside and to the unloading plate will complicate this relation and limit the voltage amplification to about 50x, as shown in fig. 17; this still means that during unloading the backside potential can reach a value of up to 1000 V. This is well above the Paschen minimum of ~275 V for $H_2$, implying a risk of electrostatic breakdown and discharges.

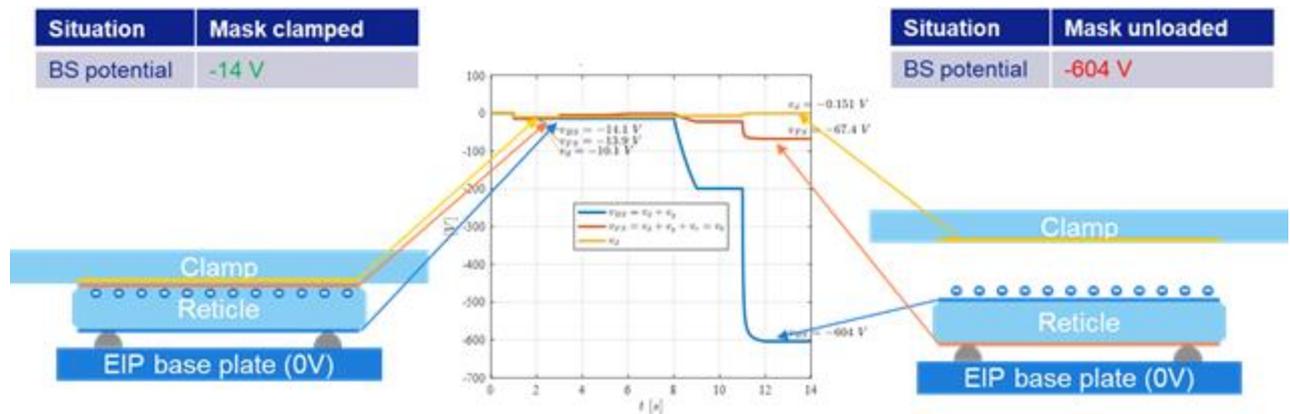

**Figure 17: Example of backside (BS) potential amplification from ~14 V to ~600 V during reticle unload (blue line), caused by increasing gap between reticle and clamp (stepwise from t=8 sec on in this capacitive model); also shown is the induced frontside voltage (red line) and the induced clamp voltage (yellow line)[49].**

The high level of reticle charging has been confirmed by electrostatic voltage measurements (ESVM), using dual Trek PD15035 555P-style probes with a 6000B-6 sensor (in combination with a modified reticle pod to allow simultaneous access of the probes to reticle front and backside), directly after unloading the reticle from the scanner. Fig. 18 shows comparison of reticle backside voltage measured after full reticle cycle through EUV machine with and without EUV exposures. Test reticles exposed to EUV confirm the high voltage of ~600 V, while reticles that were not exposed to EUV remained neutral. Optical microscope inspection of these test reticles indicated cosmetic damage of reticle backside coating after EUV exposures, which could be traced back to imperfections in the coating edges on the test reticles used, but also showed a clear sensitivity to backside discharges from these high backside voltages.

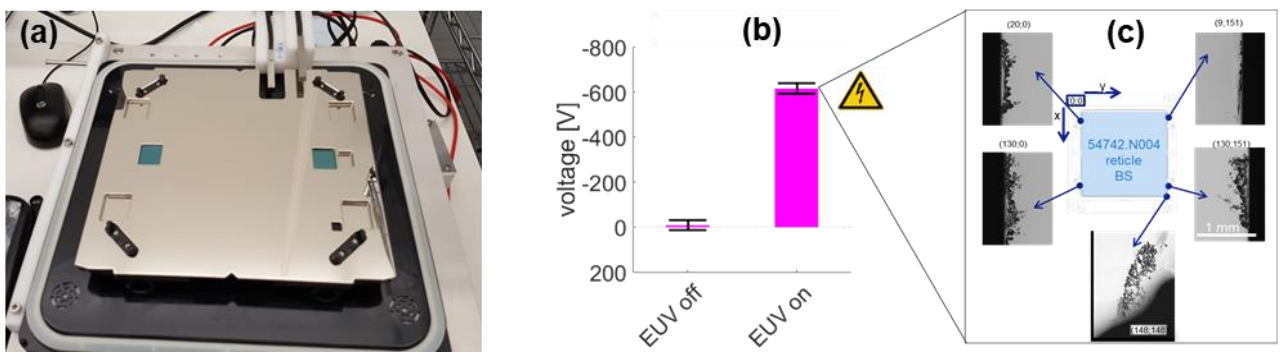

Figure 18: Left (a): modified reticle pod for ESVM measurements; Middle (b): ESVM measurement of high voltage on reticle backside when reticle has been exposed to EUV; Right (c): Observation of ESD damage on a test reticle.

Besides this backside discharge risk (which might in practice be acceptable, since the backside is not as critical as the imaging frontside of the reticle), the increase of backside voltage during unload also induces a frontside voltage in order of 70 V by their capacitive coupling (red line in fig. 17), which might not be so high as to cause concerns for discharges to the critical reticle frontside surface, but is a concern for particle attraction to the reticle, as demonstrated by Amemiya (see fig. 19)[50].

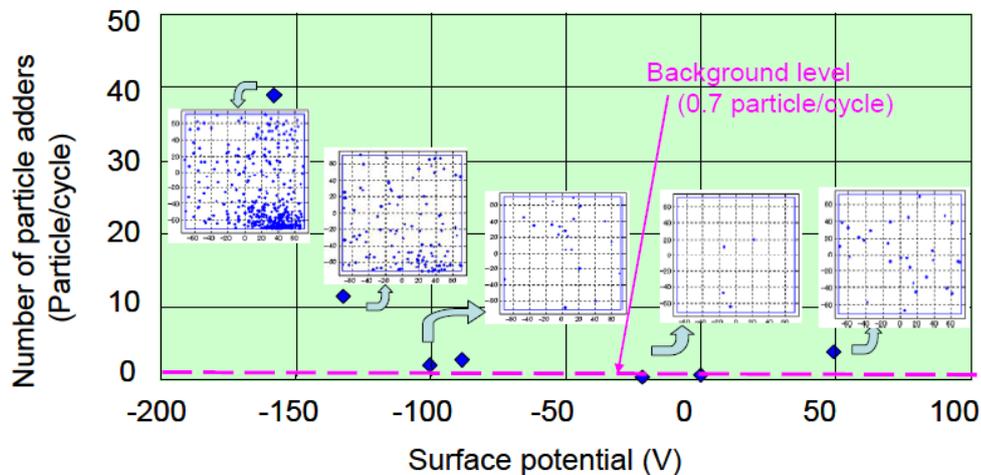

Figure 19: Particle pick-up as a function of reticle surface potential; from Amemiya[50].

Even while the EUV reticle pod is designed with electrostatics in mind (e.g. metal inner body to prevent ESD risk as present in DUV pod[51]), the pod does not fully resolve this: frontside grounding has to be soft to prevent particle generation from hard grounding contact, so will make poor electrical contact through the insulating top oxide of the reticle frontside coating for low voltages[52]. Although the backside pod cover itself is grounded, this does not make grounding contact to the reticle backside within the scanner vacuum system or the internal reticle library, but only makes contact when the EUV pod is locked at the load port to be removed from the scanner. It should be considered that this grounding is by soft contact to a potentially oxidized backside coating, so this contact might be poor in practice, and should not be relied upon for backside de-charging.

The issue of backside voltage excursions during unload can be remedied in two ways: 1) by creating a (negative) offset in the clamping scheme to shift the equilibrium of the EUV-plasma charging to (near) zero during the exposures, or 2) by supplying free charge carriers during the unload sequence to dynamically reduce the charge on the reticle as the voltage builds up.

A negative clamping offset has indeed been observed to result in lower backside voltage, with near-zero backside voltages during unload being achieved for an offset of roughly -25 V, as shown in fig. 20, using ESVM. However, as can also be seen in fig. 21, for reasons of reticle chamfer and coating tolerances this offset would need to be calibrated per reticle to guarantee sufficiently low voltage at unload. The observed limit at ~-800V is most likely an artefact of the ex-situ ESVM measurements: ESVM can only be done outside of the scanner, after fully unloading the reticle, and voltages above ~800V are expected to result in discharges during the unloading of the reticle to ambient conditions.

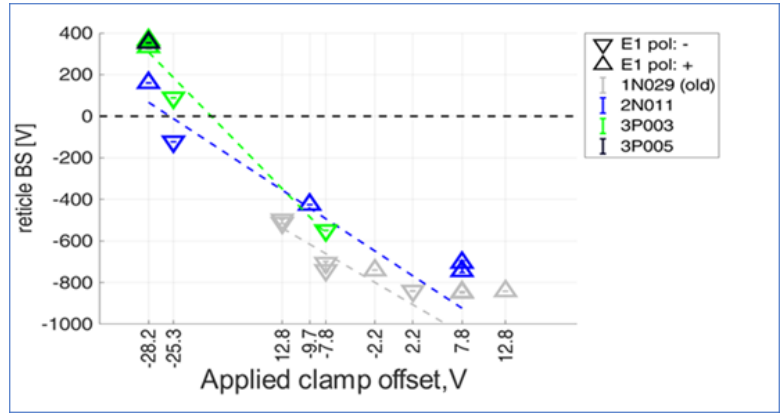

Figure 20: Reticle backside voltage as function of clamping offset; the colors denote different specimens of test reticles from two different suppliers.

An alternative solution is dynamic charge compensation during the unload sequence by creating a supply of free charge carriers; this will reduce the charge on the reticle as the voltage builds up and thus will maintain acceptably low voltage levels throughout the unloading sequence, to prevent any risk of discharge.

This could be achieved by a miniature plasma generator, such as proposed and developed by Hopwood[53]; though such a device has been shown to work for hydrogen[54], the additional hardware is hard to retrofit into the existing scanner modules and reliably igniting the hydrogen plasma at ~5 Pa is still considered a challenge. Fig. 21 shows a prototype demonstration of such a device.

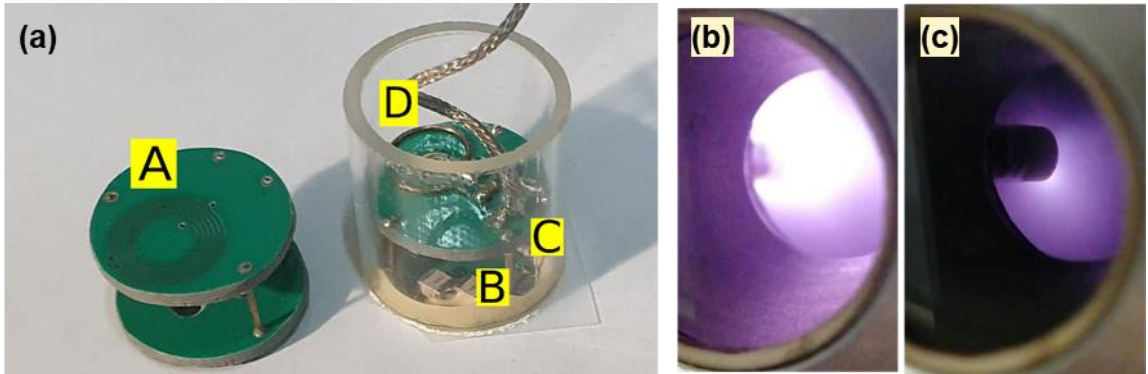

Figure 21: Left (a): prototype of miniature inductively-coupled plasma generator, with (A) inductive coil, (B) high-Q capacitors, (C) impedance matching and (D) RF current monitoring coil (only for prototype). The diameter of the assembly is approximately 2 cm. Middle (b): ignition of the discharge at 3Pa hydrogen. Right (c): stable operation at 2W RF power and 3Pa hydrogen. Images courtesy of ISAN.

Dynamic charge compensation can also be achieved by turning on the EUV-induced plasma during the unload sequence, which has been termed "EUV@unload"[55]. During the unload sequence the reticle is moved to a position next to the EUV beam, is placed onto a baseplate and subsequently lowered from the clamp. As the gap between reticle and clamp increases, the capacitance drops and the negative backside voltage builds up, attracting the ions from the EUV-plasma; simultaneously the opening gap allows the ions to reach the backside coating more easily to reduce the net charge. Even though the reticle is moved several cm's away from the EUV beam during unload, the ions are pulled towards the high negative potential on the reticle backside which develops as the reticle is moved away from the clamp. It should be noted that volume recombination is very low at this low ionization degree and pressure, so ions can travel a long distance if the directional motion in the electric field is stronger than diffusion to the walls. This is shown in fig. 22. Full scanner tests have confirmed that the EUV-induced plasma density is sufficient to counter the voltage amplification effectively, without delays or slowdowns in the unload sequence. Besides being relatively insensitive to reticle tolerances, this also has key benefits in using pre-existing hardware and having no ignition threshold. Therefore, this scheme has been chosen as baseline for the current generation of EUV-scanners.

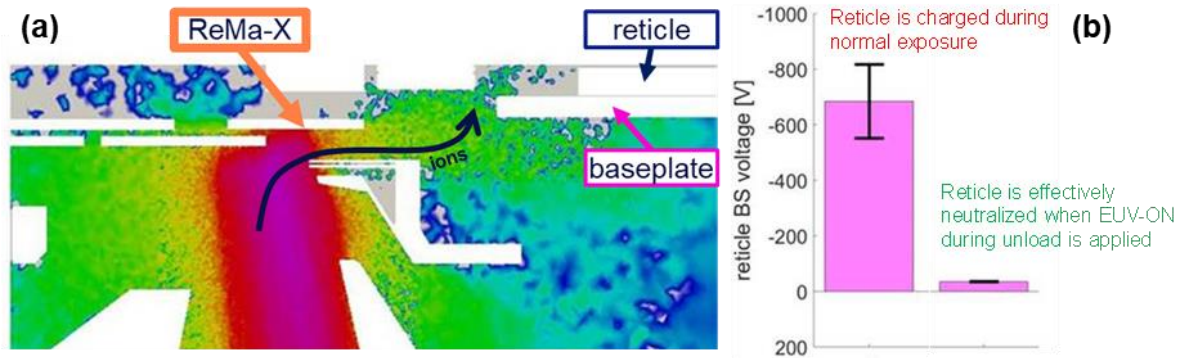

Figure 22: EUV-plasma now neutralizes reticle backside during reticle unload, even with the EUV beam several cm's away. Left (a): reticle location with respect to EUV beam and modeled ion densities; Right (b): reticle backside voltages measured by ESVM, without and with EUV on during unload.

Recent customer data has shown that EUV@unload significantly suppresses defectivity associated with electrostatic pick-up from the reticle pod baseplate, such as carbon-based fall-on particles, without deterioration of other particle types. Also, in-house testing on system (which had a known grounding issues) showed the effectiveness of EUV@unload to mitigate the resulting defectivity issue, as shown in fig. 23.

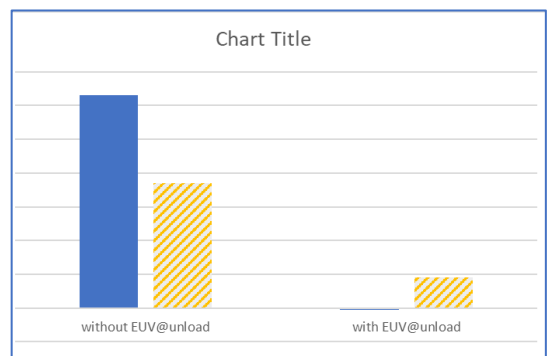

Figure 23: Effective mitigation of ESD-related particles (solid blue) by EUV@unload; ~4x improvement in non-ESD, or fall-on, particles (orange striped) cannot be attributed to EUV@unload but is likely due to flushing in between the two measurements.

Although current performance of the EUV@unload scheme is satisfactory, further improvements or accelerations are currently being investigated. One option could be to combine EUV@unload with clamp electrode biasing as outlined above, or alternatively to apply a negative bias voltage to the clamp during unloading to attract more ions from the EUV-plasma. For the long term, it is recommended to investigate grounding of both reticle surfaces during scanning; one option could be to ground the backside via hard electrical contact to the clamp and to create electrical connection between backside and frontside coatings.

5. **Particle Transport and reticle Protection**

Another electrostatic aspect of EUV is that, as shown above, during the EUV pulse the floating reticle surface will instantaneously charge positively by photoelectric effect to ~20-40 V, and subsequently will be neutralized within ~5 μs to zero volt by charge compensation from the EUV-induced plasma. This process repeats after 20 μs with the next pulse. So on average the reticle will be charged ~1-2 V positively with respect to the surrounding grounded surfaces. As the reticle top layers (Ru cap and Ta absorber) are conductive and continuous, all of the reticle will take on this average positive potential, also the (large) part of the reticle that is away from the actual EUV-beam. Away from the EUV-beam, plasma density is too low to effectively shield the resulting electric field between reticle and the nearby grounded reticle masking blades, which can be in order of >100 V/m given the small gap.

Free particles are preferentially charged negatively in and around the EUV-beam[56], although there might be transient phase of positive charging by photo-electric effect[57]. This results in an attractive electric force between reticle and particle. Charging of free particles inside the EUV beam will show transients with the EUV pulses, first charging positively by photo-electric effect, then negatively due to the higher mobility of the plasma electrons, and subsequently (partly) neutralizing due to the ions, as described by Orbital Motion Limited (OML) theory[58]. Assuming thermal equilibrium, OML provides the steady state potential of the particle $\phi_p$ when the electron and ion fluxes are balanced:

$$\exp\left(\frac{e\phi_p}{k_B T_e}\right) = \sqrt{\frac{m_e T_i}{m_i T_e}}\left(1 - \frac{e\phi_p}{k_B T_i}\right) \quad (5)$$

With $m_e, m_i$ and $T_e, T_i$ the masses and temperatures of the electrons and ions respectively. With the ions close to room temperature, the particle potential is mainly determined by the electron temperature. Approximating the particle by a sphere, the particle charge $Q_p$ follows from the potential via the capacitance of a sphere, and scales linearly with the particle diameter $d_p$[59].

$$Q_p = 2\pi\epsilon_0 \cdot d_p \cdot \phi_p \quad (6)$$

In the transient EUV-induced plasma, no analytical equations exist for the potential or charge of a free particle, and PIC modeling is used to determine the evolution of particle charge over time. PIC simulations of the EUV-induced plasma in the region below the reticle show that micron-sized particles get a short positive charge, after which they reach an equilibrium between electron and ion currents, as illustrated in fig. 24. For submicron particles, neutralization will typically take longer than the pulse interval, and negative particle charge will build up over multiple pulses until an equilibrium is reached between the photo-ionization and the electron currents, as illustrated in fig. 25. For an electron temperature of ~0.5 eV, the equilibrium particle charge is predicted to be roughly $2d_p$ [e], with $d_p$ the particle diameter.

Free particles next to the EUV beam will not experience the initial photo-electric effect, so will charge negatively by the more mobile electrons and will reach an equilibrium between electron and ion collection currents; this results in a similar steady-state value negative charge as for particles within the beam, or somewhat more negative for submicron particles. Further away from the beam, plasma density will drop and the charging will be much slower, but will still result in an average negative charge (again due to the more mobile electrons).

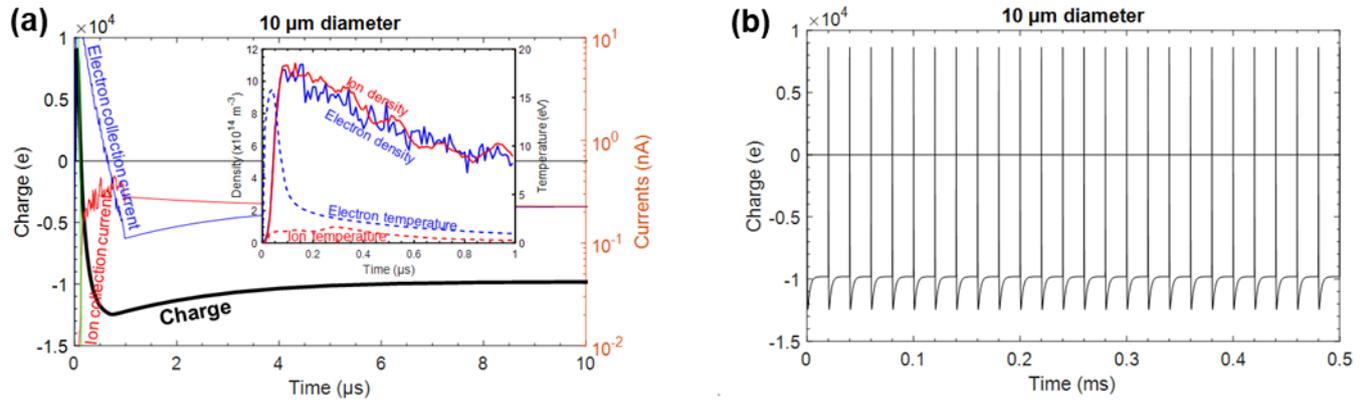

Figure 24: Left (a): PIC model of charging of a 10 μm particle in the EUV beam, showing fast transient positive photo-charging and subsequent negative charging (to ~1 e per nm); quasi steady-state charge (balanced electron and ion currents) is achieved within ~5 μs. The insert shows the electron and ion temperatures. Right (b): repeating charging pattern over multiple pulses.

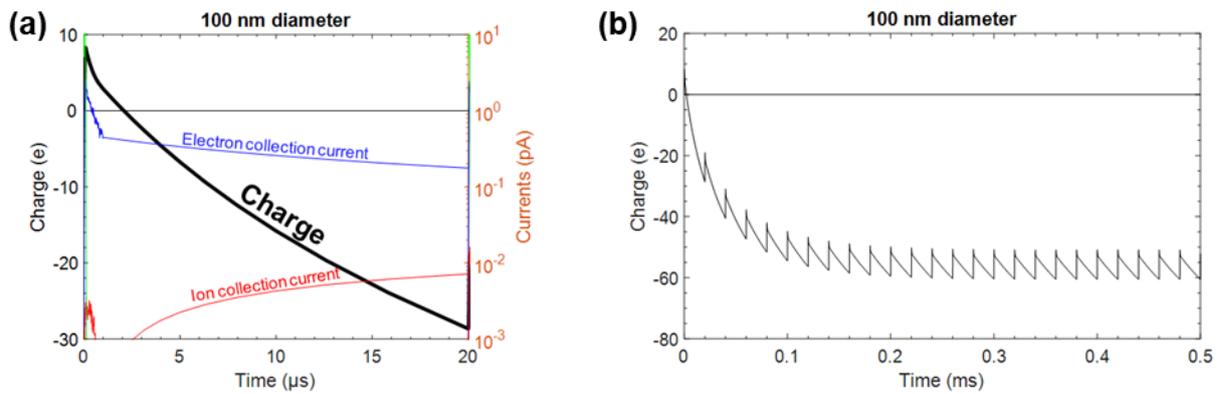

Figure 25: Left (a): PIC model of charging of a 100 nm particle in the EUV beam, showing fast transient positive photo-charging and subsequent negative charging. Right (b): increasing charge for over multiple pulses for 100 nm particle.

Extending the PIC simulations for particle charge with dynamic reticle potential and resulting electric fields yields an electric force on the particle near the reticle surface. Comparing the resulting electric force against the other forces that might work on a free-floating particle (gravitational force, neutral and ion drag forces, and for completeness, thermophoretic force (driven by temperature differences between the irradiated reticle surface and reticle-facing masking blades)), it is clear that the dominant forces for submicron particles are the electric force and the neutral drag force, as illustrated by fig. 26.

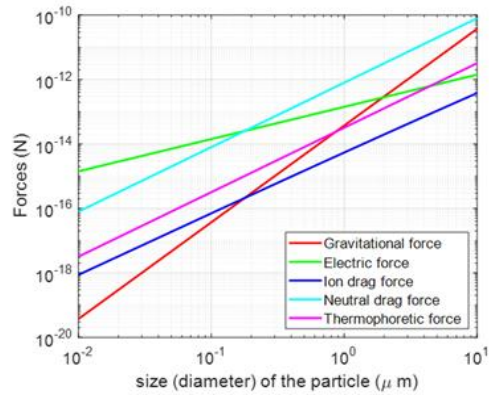

Figure 26: Volume force estimates for submicron particles in the gap between reticle and grounded reticle masking blades, for 250W Source and 5 Pa $H_2$.

Combining the vector force fields of electric and neutral drag forces, particle trajectories can be calculated, as shown in fig. 27. This allows to design the local flows and pressures such that no particles larger than a given critical size (50 nm in this example) will reach the reticle frontside surface. The most effective optimization parameter is pressure: increasing pressure will increase neutral drag force[60] and reduce the electric attraction force (as outlined above), but will come at expense of EUV transmission. Increasing flow in itself will increase neutral drag force and does not affect the electric force, but in practice flow is not independent of pressure.

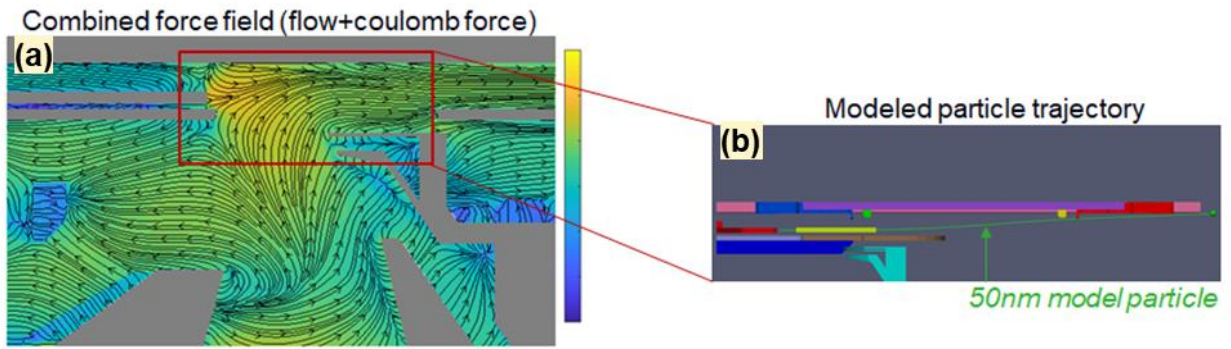

Figure 27: Left (a): Particle-in-Cell simulations of the force map for a 50-nm diameter particle in the reticle zone, just after the EUV pulse (180 ns). More yellow means higher force (blue means lower); arrows indicate local direction of force. Right (b): modeled trajectory for a 50 nm test particle

6. **Conclusion**

Understanding of the specific nature of the EUV-induced scanner plasma and the interaction of this plasma with surfaces and particles has improved significantly in the past years. This has allowed scanner design optimizations as well as targeted improvements in manufacturing and cleaning processes, for both plasma and related electrostatics aspects, to continue to drive down the corresponding contributors behind defectivity. Analysis of plasma-assisted discharges has resulted in design guidelines for allowed voltages and pressures at various distances from the EUV-beam, to prevent discharges and improve the robustness of high-voltage electronics. Prevention of high reticle charging during unloading and handling, with the associated risk of discharges, removes a potential source of particles. These measures have brought particle contamination control of the EUV-scanner to a regime where customers have the freedom to operate without protective pellicle for high-volume manufacturing. For future EUV lithography systems, continued improvements in plasma models and understanding will ensure plasma-aware designs that will be compatible at increasing source powers and reducing critical particle sizes.

## 7. acknowledgements

The authors wish to thank the ASML Research and Development teams for Scanner Plasma and Defectivity for scanner testing, fruitful discussions and general support. Also, we would wish to thank Ronald van der Wilk, Alex Yanson, Frank van Lier and Jeroen van Duivenbode for the electrostatic measurements and electrostatic reticle model; and Edgar Osorio Oliveros and Vladimir Krivtsun, Maxim Spiridonov, Slava Medvedev of ISAN for the miniature ICP work. And finally, we would like to thank Pavel Krainov, Bogdan Lakatosh and ISAN for the PIC simulations.

*Biography*

**Mark van de Kerkhof** began his career at ODME, developing then-novel DVD mastering, and later worked on Deep-UV and Immersion technologies for Blu-Ray. In 1999 he joined ASML as senior designer, working on miscellaneous projects for both DUV and EUV scanners, and was responsible for the technical definition and integration of the NXE:3400B EUV scanner as Product System Engineer. He is currently responsible for EUV Scanner Plasma Technology. He co-authored more than 30 scientific papers and holds more than 70 USA patents.

**Caption List**

Figure 1: Basic principle of an EUV scanner; the object on a reticle (or mask) is illuminated and imaged onto a portion of a wafer by the Projection Optics Box (POB) while scanning; after which the wafer is moved to a new position and the process is repeated (source: ASML).
Figure 2: Basic processes of EUV-induced hydrogen plasma.
Figure 3: Left: Build-up, steady-state and decay of pulsed EUV-induced plasma, as measured by Retarding Field Energy Analyzer (RFEA) in scanner-like test-stand directly after LPP Source exit. Right: zoom-in on first two pulses; also visible is the minor plasma formation due to the pre-pulse.
Figure 4: Schematic of reticle zone, showing EUV beam region (A), floating reticle surface (B), grounded reticle masking blades (C). Left (a): during isolated EUV pulse the electrons from the EUV beam penetrate through the slits. Right (b): plasma accumulation beyond the beam confines over multiple pulses.
Figure 5: Left (a): Paschen curve for H2, showing good match between analytical expression and measurements. Right (b): Modified breakdown curves in presence of a surface (insulating epoxy, with different nanoparticle additions), showing a more gentle slope at low p.d-values (reproduced from Li).
Figure 6: Left (a) and middle (b): gradual voltage ramp at 2 V/s until breakdown for 10 and 40 Pa; blue line is voltage to frontside, red line is frontside voltage as measured by ESVM. Right (c): summary of the breakdown voltages for different p.d values (no discharge was observed at 5 Pa for up to 800 V maximum).
Figure 7: Simplified 2D model geometry. Depicted are the cylindrical tube with a dielectric wall (gray), electrodes (orange) and plasma-filled region (pink). Neutral H2 gas is represented with green dots.
Figure 8: Simulated plasma dynamics for 10 Pa (a,b) and 1 Pa (c,d), showing electron and ion densities (left and right side of each image) and voltage contour lines, for specific time stamps. Anode potential was set to 500 V. The snapshots for 10 Pa show the transition to breakdown at ~3 ms, while the snapshots for 1 Pa show the more or less stable anode glow.
Figure 9: Model showing avalanche and breakdown at 10 Pa, with sharply peaked increase in current (green line); and oscillating glow discharge at 1 Pa (purple line).
Figure 10: Experimental setup for plasma-assisted discharge: 1 – Cesar 1312 RF generator; 2 – stainless steel chamber; 3 – hydrogen valve; 4 - hydrogen plasma; 5 – glass tube; 6 - micrometer translation stage; 7 – insulated PTFE rod with a connector; 8 – power supply/multimeter; 9 – stainless steel electrode; 10 – pressure gauge; 11 – turbo-pump; 12 – rotary pump.
Figure 11: Left (a): example of plasma-assisted discharge and current when ramping up voltage between electrodes while plasma switched on in neighboring chamber (using 60W plasma, 5 Pa, 10 cm distance). Right (b): examples of significant reduction of breakdown voltage when plasma is switched on (10W plasma; 5 cm distance; 1, 2 and 5 Pa); both with and without plasma, the observed breakdown voltages at low pressures are all significantly reduced with respect to classical Paschen prediction of >105 V (see fig. 6), which is attributed to the glass tube wall surfaces.
Figure 12: Sketch of reticle cross-section, showing conductive coating stacks on frontside and backside; adapted from McLellan
Figure 13: PIC simulation of electron density (blue line) and reticle frontside surface potential (red line), showing accumulation of plasma over pulses but no build-up of surface potential.
Figure 14: Left (a): basics of charging mechanism of reticle backside; Top right (b): energy spectrum of electrons reaching reticle edge modelled by 3D PIC code. Bottom right (c): energetic secondary electrons charge reticle backside, while ions are carried by field and momentum towards nearest wall resulting in a net negative charge on reticle.
Figure 15: PIC simulations of backside voltage dependence on time and pressure, showing >2x reduction in voltage for 1.5x higher pressure.
Figure 16: Illustration of reticle unloading sequence. While clamped the separation between reticle backside and clamp electrode is in order of a few micrometers, resulting in strong capacitive coupling; during unload the separation is increased to ~1 cm, reducing the capacitance by several orders of magnitude.

Figure 17: Example of backside (BS) potential amplification from ~14 V to ~600 V during reticle unload (blue line), caused by increasing gap between reticle and clamp (stepwise from t=8 sec on in this capacitive model); also shown is the induced frontside voltage (red line) and the induced clamp voltage (yellow line).

Figure 18: Left (a): modified reticle pod for ESVM measurements; Middle (b): ESVM measurement of high voltage on reticle backside when reticle has been exposed to EUV; Right (c): Observation of ESD damage on a test reticle.

Figure 19: Particle pick-up as a function of reticle surface potential; from Amemiya50.

Figure 20: Reticle backside voltage as function of clamping offset; the colors denote different specimens of test reticles from two different suppliers.

Figure 21: Left (a): prototype of miniature inductively-coupled plasma generator, with (A) inductive coil, (B) high-Q capacitors, (C) impedance matching and (D) RF current monitoring coil (only for prototype). The diameter of the assembly is approximately 2 cm. Middle (b): ignition of the discharge at 3Pa hydrogen. Right (c): stable operation at 2W RF power and 3Pa hydrogen. Images courtesy of ISAN.

Figure 22: EUV-plasma now neutralizes reticle backside during reticle unload, even with the EUV beam several cm's away. Left (a): reticle location with respect to EUV beam and modeled ion densities; Right (b): reticle backside voltages measured by ESVM, without and with EUV on during unload.

Figure 23: Effective mitigation of ESD-related particles (solid blue) by EUV@unload; ~4x improvement in non-ESD, or fall-on, particles (orange striped) cannot be attributed to EUV@unload but is likely due to flushing in between the two measurements.

Figure 24: Left (a): PIC model of charging of a 10 mm particle in the EUV beam, showing fast transient positive photo-charging and subsequent negative charging (to ~1 e per nm); quasi steady-state charge (balanced electron and ion currents) is achieved within ~5 ms. The insert shows the electron and ion temperatures. Right (b): repeating charging pattern over multiple pulses.

Figure 25: Left (a): PIC model of charging of a 100 nm particle in the EUV beam, showing fast transient positive photo-charging and subsequent negative charging. Right (b): increasing charge for over multiple pulses for 100 nm particle.

Figure 26: Volume force estimates for submicron particles in the gap between reticle and grounded reticle masking blades, for 250W Source and 5 Pa H2.

Figure 27: Left (a): Particle-in-Cell simulations of the force map for a 50-nm diameter particle in the reticle zone, just after the EUV pulse (180 ns). More yellow means higher force (blue means lower); arrows indicate local direction of force. Right (b): modeled trajectory for a 50 nm test particle